\documentclass{aa}  

\usepackage{graphicx}
\usepackage{amsmath}
\usepackage{indentfirst}
\usepackage{txfonts}

\def\myr {$\rm M_\odot\,yr^{-1} \,$}
\def\ergs {$\rm erg\,s^{-1}\,$}
\def\ergscm {$\rm erg\,s^{-1}\,\,cm^{-2} \,$}
\def\ergscma {$\rm erg\,s^{-1}\,\,cm^{-2}\,\AA^{-1} \,$}

\begin{document}

   \title{{\it K} band SINFONI spectra of two $z \sim 5$ SMGs: upper limits to the un-obscured star formation from [O\,{\sc ii}] optical emission line searches}

   \author{Guilherme S. Couto\inst{1,2} 
	   \and Luis Colina\inst{2,3} 
	   \and Javier Piqueras L{\'o}pez\inst{2}\and \\
	   Thaisa Storchi-Bergmann\inst{1}
	   \and Santiago Arribas\inst{2,3}}

   \institute{Universidade Federal do Rio Grande do Sul, IF, CP 15051, Porto Alegre 91501-970, RS, Brazil \\
              \email{gcouto@if.ufrgs.br}\\
	      \and
	      Centro de Astrobiolog\'{i}a (INTA–CSIC), Ctra de Torrej\'{o}n a Ajalvir, km 4, 28850 Torrej\'{o}n de Ardoz, Madrid, Spain\\
	      \and
	      ASTRO-UAM, UAM, Unidad Asociada CSIC, Spain\\}

    \titlerunning{{\it K} band spectra of two $z \sim 5$ SMGs}
    \authorrunning{Guilherme S. Couto et al.}

   \date{Received -; accepted -}

\abstract{We present deep SINFONI {\it K} band integral field spectra of two submillimeter (SMG) galaxy systems: BR 1202-0725 and J1000+0234, at $z=4.69$ and $4.55$ respectively. Spectra extracted for each object in the two systems do not show any signature of the [O\,{\sc ii}]$\lambda\lambda$3726,29\AA\,emission-lines, placing upper flux limits of $3.9$ and $2.5 \times 10^{-18}\,$\ergscm for BR 1202-0725 and J1000+0234, respectively. Using the relation between the star formation rate (SFR) and the luminosity of the [O\,{\sc ii}] doublet from \citet{kennicutt98}, we estimate unobscured SFR upper limits of $\sim$ $10-15\,$\myr and $\sim$ $30-40\,$\myr for the objects of the two systems, respectively. For the SMGs, these values are at least two orders of magnitude lower than those derived from SED and IR luminosities. The differences on the SFR values would correspond to internal extinction of, at least, $3.4 - 4.9$ and $2.1 - 3.6$ mag in the visual for BR 1202-0725 and J1000+0234 SMGs, respectively. The upper limit for the [O\,{\sc ii}]-derived SFR in one of the LAEs (Ly$\alpha2$) in the BR1202-0725 system is at least one order of magnitude lower than the previous SFR derived from infrared tracers, while both estimates are in good agreement for Ly$\alpha$1. The lower limits to the internal extinction in these two Lyman-alpha emitters (LAEs) are $0.6$ mag and $1.3$ mag, respectively. No evidence for the previously claimed \citep{ohta00} [O\,{\sc ii}] emission associated with Ly$\alpha$1 is identified in our data, implying that residuals of the K-band sky emission lines after subtraction in medium-band imaging data could provide the adequate flux.}


   \maketitle


\section{Introduction}

\begin{figure*}
\centering
\includegraphics[width=\textwidth]{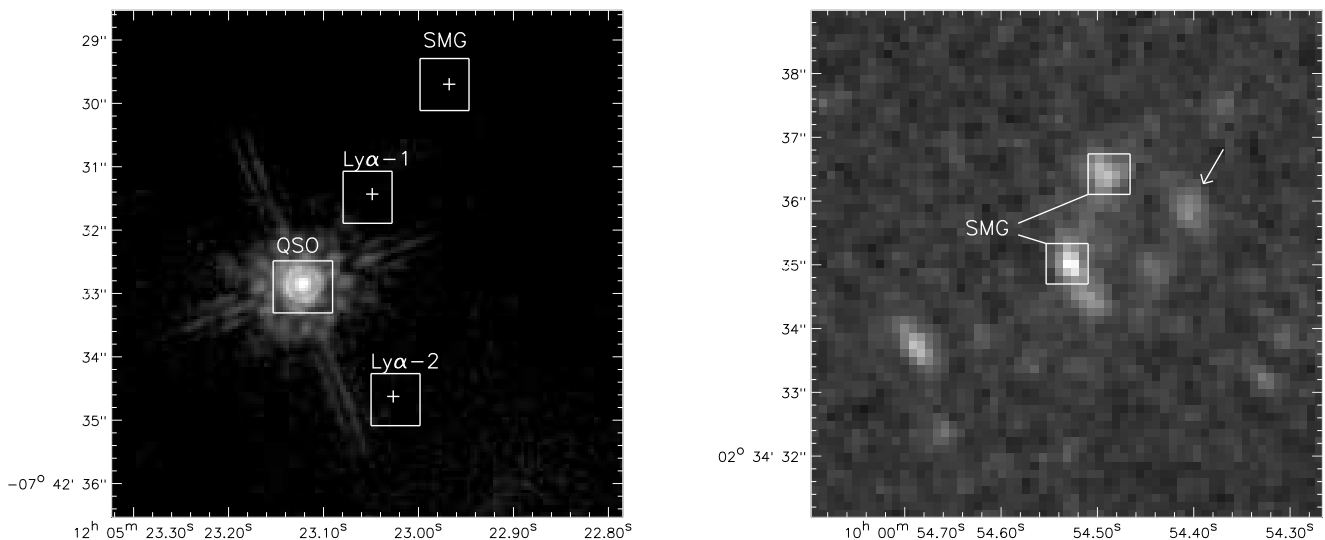}
\caption{Left panel: {\it HST-NICMOS 2} image of the BR1202-0725 system, using the F160W filter, covering the SINFONI FoV of $8\arcsec \times 8\arcsec$. White crosses mark the positions of the SMG, quasar, Ly$\alpha1$ and Ly$\alpha2$ as labeled, according to \citet{carilli13}. White squares show the regions where spectra were extracted from the SINFONI datacube. Right panel: {\it HST-WFC3} image of J1000+0234, using the IR channel and F160W filter. White squares show the regions where SMG spectra were extracted from the SINFONI datacube, as identified. The white arrow points to a foreground object.}
\label{hst}
\end{figure*}

Recent studies of galaxies detected at millimeter and submillimeter wavelengths have largely increased our understanding of the formation and evolution of massive galaxies when the Universe was 1-3 Gyrs old. Cosmological simulations indicate that massive galaxies can form at high-$z$ via gas-rich mergers, triggering extreme events such as intense star formation and simultaneously growing of supermassive black holes (SMBHs) close to their maximum (Eddington) accretion rates \citep{li07,netzer14}. Submillimeter galaxies (SMGs) play a key role in this scenario, since they represent examples of starburst galaxies (above main sequence) in the distant universe. At such high redshifts ($z\sim5$), a large fraction of the star formation activity is enshrouded in dust, most of the bolometric luminosity is radiated into the far-infrared (FIR, $40-500\,\mu$m) and submillimeter wavelengths and therefore the luminosity of galaxies is proportional to the star formation rate (SFR). SMGs observed with {\it Herschel} and ALMA \citep{mor12,vieira13,netzer14} have displayed extremely high infrared luminosities ($L_{FIR} \geq 10^{13} L_\odot$), which implies SFRs of $> 10^3 M_\odot\,$year$^{-1}$. These rates indicate that the bulk of the star formation in these galaxies have timescales of only $\leqslant 100\,$Myr. 

The structure and physical mechanisms at work in these objects are largely unknown. Different scenarios have been proposed that imply the removal of the available gas due to an active galactic nucleus (AGN) driven wind in luminous quasars or to supernovae and stellar winds in extreme starbursts, leaving behind a compact remnant \citep{sanders88,hopkins08,wuyts10}. However, evidence for significant outflow rates at high-$z$ are very limited and have only recently been achieved. One example is the highly magnified galaxy at $z = 4.92$ behind the lensing cluster MS 1358+62 \citep{swinbank09}, in which deep [O\,{\sc ii}]$\lambda\lambda$3726,29\AA\, spectroscopy indicates the presence of a young outflow ($< 15\,$Myr) and a SFR of $42$ \myr.

In this work, we present deep SINFONI--IFS data of two $z \sim 5$ SMGs, BR 1202-0725 and COSMOS J100054+023436. BR 1202-0725 ($z = 4.69$) is a system composed by a dusty, luminous starburst (the SMG itself), an optically luminous QSO and two Ly$\alpha$ emitting extented regions, hereafter Ly$\alpha1$ and Ly$\alpha2$ \citep{omont96,carilli02}. The QSO and the SMG are separated by $3\farcs8$, with Ly$\alpha1$ between them ($2\farcs3\,$ north-west of the QSO), a clear signature of interaction, while Ly$\alpha2$ lies $2\farcs7\,$ south-west of the QSO (see left panel of Figure \ref{hst}). Recent observations of the system using ALMA detects narrow [C\,{\sc ii}]$\lambda158\,\mu$m emission in all four sources \citep{wagg12,carilli13}. Both SMG and QSO have high FIR luminosity ($L_{FIR} > 10^{13} L_\odot$) indicating SFRs of the order of $10^3\,$\myr or above \citep{iono06,carniani13}, while both Ly$\alpha$ emitting regions, Ly$\alpha1$ and Ly$\alpha2$, appear to be forming stars, added together, at a rate of $19$ \myr \citep{williams14}, derived from the [C\,{\sc ii}] $158 \mu$m emission-line luminosities. \citet{carilli13} concluded that the proximity of a luminous quasar is unlikely to be the source of the ionized nebula in Ly$\alpha1$, and speculate that a optically thick torus is shielding the radiation towards the Lyman-alpha emitters (LAEs, i.e. Ly$\alpha$1 and Ly$\alpha$2). Gas outflows related to the QSO are under debate, with some evidence based on the detection of a broad [C\,{\sc ii}] secondary component \citep{carilli13}, not confirmed by similar subsequent studies \citep{carniani13}. \citet{ohta00} report detection of [O\,{\sc ii}] doublet emission of the Ly$\alpha1\,$ galaxy based on narrow-band imaging. 

COSMOS J100054+023436 (hereafter J1000+0234, $z = 4.55$) is a SMG dominated by a starburst forming stars at a rate of $> 1000\,$\myr with a young age of $2-8\,$ Myr, estimated using infrared and radio measurements \citep{capak08}. This SMG presents multiple components observed in the Ly$\alpha$ emission-line, distributed along a region of $\approx 3''$ in size, and residing at the same redshift. This galaxy presents different morphology at different wavelengths, from the UV (Ly$\alpha$) to the near-infrared (rest-frame optical with [O\,{\sc ii}] doublet), likely indicating the presence of spatially resolved stellar populations on arcsec scales \citep{capak08}. Detection of a broad CO($4-3$) emission line indicates the presence of a large amount of molecular gas with an estimated total mass of $2.6 \times 10^{10}\,\textrm{M}_\odot$. The width of the CO($4-3$) line and the highly disturbed morphology suggest this system is involved in an ongoing merger \citep{schinnerer08}.

The objective of this work is to search the [O\,{\sc ii}]$\lambda\lambda$3726,29\AA\, doublet in these two SMGs, to study the structure of the un-obscured star formation, including the presence of flows of gas other than rotation. This paper is organized as follows: in Sec. \ref{obs} we describe the observations and data reduction, in Sec. \ref{res} we present the results from the SINFONI datacubes for both SMG systems. In Sec. \ref{analys} we discuss the implications for the derived star formation rate, internal extinction in the visual and the claim of previous [O\,{\sc ii}] detection in the Ly$\alpha1$ of the BR1202-0725 system. Finally, conclusions are summarized in Sec. \ref{conc}. Thoughout this paper we assume a standard concordance cosmology ($H_0 = 70$, $\Omega_M = 0.3$, $\Omega_\Lambda = 0.7$).

\section{Observations and data reduction}
\label{obs}

The data were obtained in service mode between January and July 2015 using the near-IR integral field spectrograph SINFONI on the VLT (period 93A). The objects were observed in the {\it K} band (1.95-2.45 $\mu$m), with each exposure sliced into 32 slitlets imaged onto 64 pixels of the detector, leading to a spatial scale of $0\farcs125 \times 0\farcs250\,$pixel$^{-1}$. We used different observation strategies for the two objects. For BR1202-0725, 15 independent Observing Blocks (OB), each with an on-source + sky pattern given by the O-S-S-O-O-S sequence were executed. The total on-source exposure time is $3.7\,$h, and the field of view (FoV) is $\approx 10'' \times 8''$. For J1000+0234, 2 on-source exposures of 600\,s each were obtained in 15 OBs (except for one OB where exposures of 300\,s were taken), following an O-S-S-O pattern for on-source and sky frames. In this case, since the object is compact and smaller than the field-of-view of SINFONI, the sky exposures were obtained $3\arcsec$ east of the on-source pointing, thus also containing the object within SINFONI FoV. The resulting total on-source exposure time is then $9.7\,$h, with a total FoV of $\approx 14'' \times 8''$.

\begin{figure}
\centering
\includegraphics[width=0.5\textwidth]{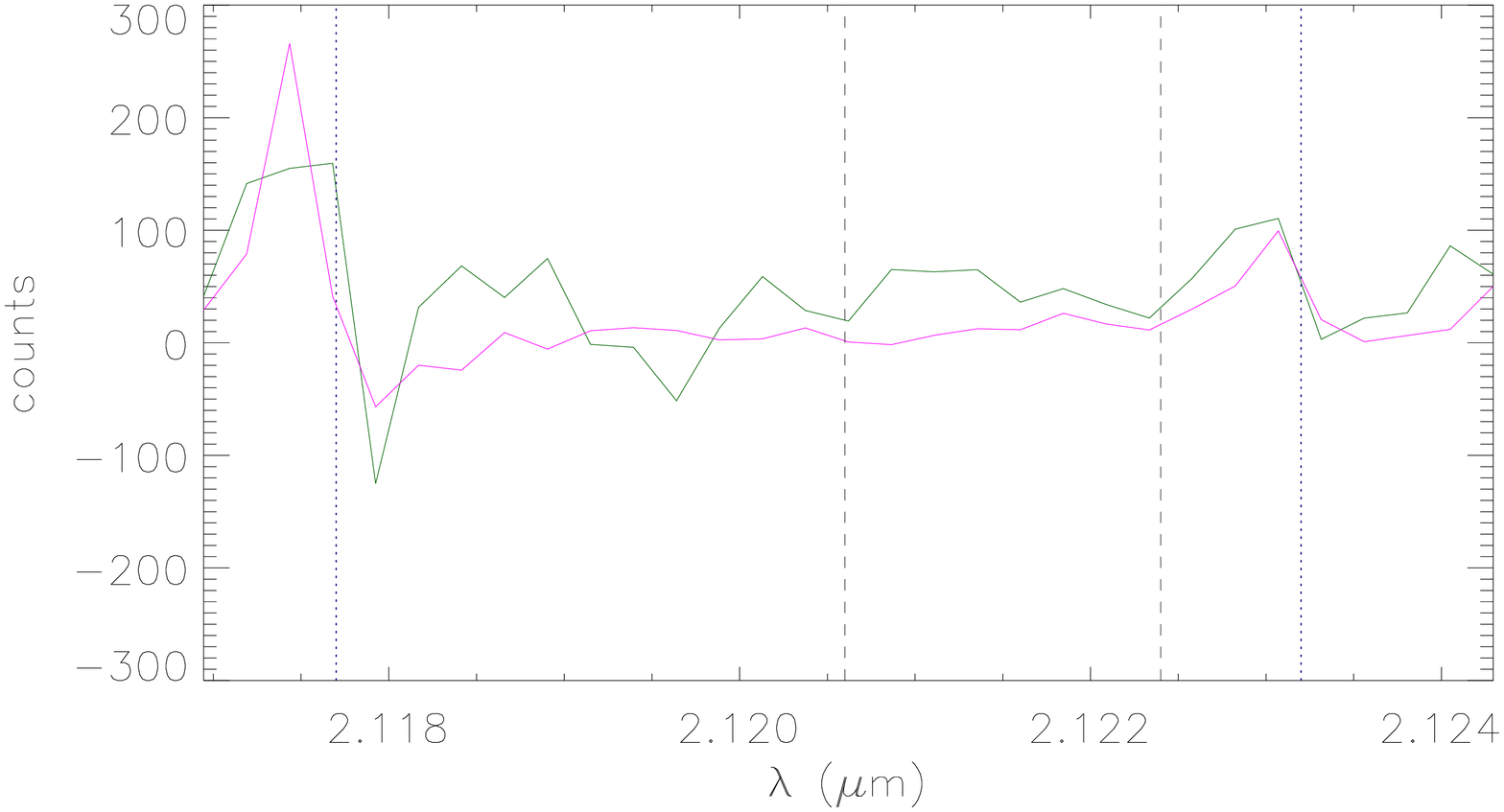}
\caption{A comparison between the standard sky subtraction applied in the data reduction previous to combining the individual datacubes (magenta line) and the sky subtraction applied after constructing the final datacube (green line), for the SMG of BR1202-0725. The latter method is the one used in this work. The wavelengths of the [O\,{\sc ii}] doublet are shown in vertical dashed black lines and dotted vertical lines indicate sky lines.}
\label{sky_sub}
\end{figure}

The spectral resolution for the {\it K} band is $R \sim 3100-3200$, and the full width at half maximum (FWHM) measured from the OH sky lines is ($6.9\pm0.3\,$\AA), with a dispersion of $2.45\,$\AA/pixel. Aside from respective sky frames exposures, a set of spectrophotometric standard stars were also observed, to correct the data for instrument response and for flux calibration. Reduction process was perfomed using the standard ESO pipeline ESOREX (version 3.10.2), and own IDL routines for flux calibration. The resulted calibration present flux uncertainties of $\approx 6\%\,$ ($1\,\sigma$) for the absolute flux, estimated as the standard deviation of the flux factors from the different calibration stars (17 and 14 standard stars observed for BR1202-0725 and J1000+0234, respectively). Individual frames were corrected for dark subtraction, flat-fielding, detector linearity, geometrical distortion and were wavelength calibrated. We have applied two different methods to address the emission due to the sky (both lines and thermal continuum). We applied the standard method of subtracting the closest sky frame to each target frame for each Observing Block (OB), and combining the individual sky subtracted cubes per OB into a single final data cube. The results are indicated by the magenta line in Fig. \ref{sky_sub} for BR1202-0725. Since the redshifted [O\,{\sc ii}] lines were expected to land in between the sky OH lines, an average sky thermal continuum emission was generated with all sky frames. This was subtracted from the final data cube with the spectrum of the object. The end result is presented with the green line in Fig. \ref{sky_sub}. Since the noise level with this method is somewhat more conservative, i.e. higher noise, we decided to use this method and the corresponding residuals to place more conservative upper limits to the [O\,{\sc ii}] flux. Cubes were then constructed from the individual frames, and the several pointings were combined to build a final mosaic, with each spectrum in the cube corresponding to an area of $0\farcs125\, \times 0\farcs125$. To shift and register all individual cubes into a single final data cube, we used the right ascension and declination keywords in the header, since the informed offsets were not matching the relative shifts as measured by the position of the QSO (relative shifts of up to $0\farcs5$ were measured in a few cases). The accuracy of this procedure is given by the spatial size of a pixel ($0\farcs125$), since the coordinates information are related to the central pixel of each individual datacube. In the case of BR1202-0725, we have also derived a final sky datacube, averaging the individual sky frames. Our seeing-limited observations have an average angular resolution of $0\farcs71\pm0\farcs26\,$ and $0\farcs59\pm0\farcs09\,$ for BR1202-0725 and J1000+0234 respectively, obtained from observations of the accompanying standard stars and assuming a gaussian profile.

\section{Results}
\label{res}

\subsection{BR1202-0725}

\begin{figure*}
\centering
\includegraphics[width=\textwidth]{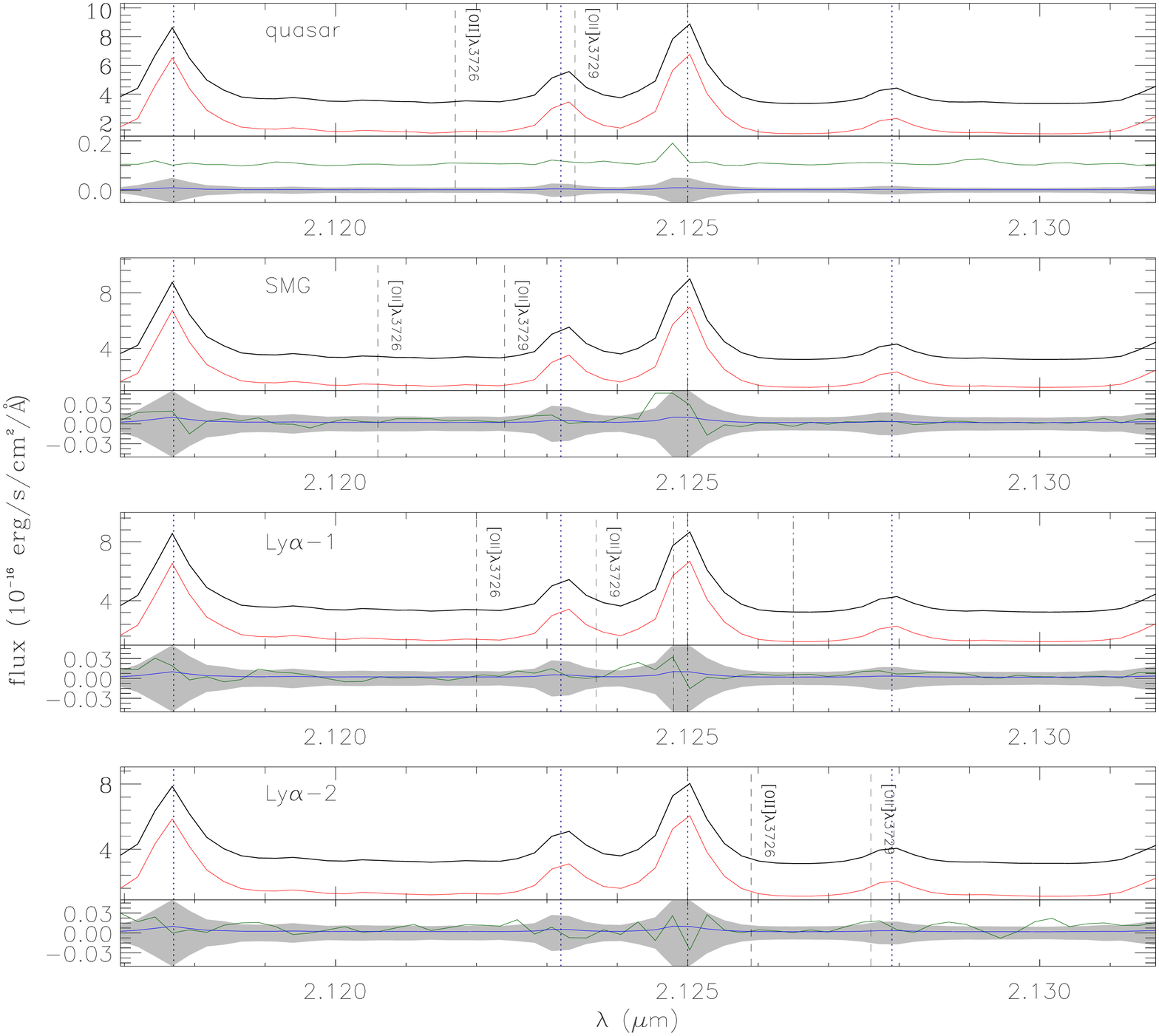}
\caption{Black lines: Integrated spectrum corresponding to the [O\,{\sc ii}] doublet range for the quasar, SMG, Ly$\alpha1$ and Ly$\alpha2$ (from top to bottom) of the BR1202-0725 system, extracted from the square apertures shown in the left panel of Fig. \ref{hst}, shifted by a constant ($2 \times 10^{-16}\,$\ergscma), added for visual purposes. Red lines: Sky spectra, which were derived as the median of the sky spectra extracted from the same apertures shown in Fig. \ref{hst}. Green lines show the residual between object and sky spectra and 1$\sigma$ uncertainty of the sky spectra are displayed in blue lines. Gray shaded areas display the 5$\sigma$ level. The wavelengths of the [O\,{\sc ii}] doublet considering [C\,{\sc ii}] redshifts \citep{carilli13} are shown in vertical dashed black lines. Dash-dotted lines show the wavelengths from Ly$\alpha$ redshifts \citep{ohyama04} for Ly$\alpha1$ (see Table \ref{redshift}). Dotted vertical lines indicate sky lines.}
\label{spec_br}
\end{figure*}

The entire system consisting of the SMG, a QSO and two Ly$\alpha$ emitters is shown in left panel of Figure \ref{hst}. The figure displays a {\it HST-NICMOS 2} image in the F160W filter (central wavelength of $1.60\,\mu$m) of the BR1202-0725 system within the same FoV of SINFONI, along with the positions of the four components of the system marked by a white cross, except for the quasar that is clearly visible. The positions were obtained from \citet{carilli13}, based on submillimeter continuum measurements, with the exception of Ly$\alpha1$, where the peak of the [C\,{\sc ii}] line emission was used. The SMG and the Ly$\alpha$ emitters are not detected in this image. White squares correspond to $0\farcs75 \times 0\farcs75$ apertures where spectra were extrated from the SINFONI data cube. The size of the apertures matches the seeing (FWHM).

The resulting extracted spectra are displayed in Fig. \ref{spec_br}. The corresponding flux limits for the two [O\,{\sc ii}] lines of the doublet are given in Table \ref{redshift} for the different components of the BR1202-0725 system assuming redshifts given by the [C\,{\sc ii}]$158\,\mu$m lines. To estimate the fluxes given in Table \ref{redshift}, we assumed that the emission lines were unresolved at the spectral resolution provided by SINFONI, and therefore considered a width of $6.9\,$\AA (FWHM). For Ly$\alpha1$ flux estimates we assume the [O\,{\sc ii}] in emission is at the redshift of the Ly$\alpha$ line. The spectra of the SMG and LAEs in the BR 1202-0725 field display no evidence of [O\,{\sc ii}] in emission above the (5$\sigma$) sky background uncertainties with upper limits $\sim$ 0.8-1.5 $\times$ 10$^{-17}\,$\ergscm for the combined flux of the doublet. The only clear features in the spectra of both the objects and are the sky lines identified by blue dotted lines. For the QSO associated with the system, there is a significant detection of the continuum but no evidence of the [O\,{\sc ii}] emission line, given the upper flux limit corresponding to $\sim$ 9 $\times$ 10$^{-17}\,$\ergscm. 

\begin{table*}
   \centering   
   \caption{\it Redshifts and measured fluxes for the objects of the BR1202-0725 and J1000+0234 systems.}
   \begin{tabular}{c c c c c} 
      \hline \hline
      Object & Emission-line used to measure $z$ & $z$ & Flux [O\,{\sc ii}]$\lambda$3726\AA $^1$ & Flux [O\,{\sc ii}]$\lambda$3729\AA $^1$ \\
      \hline
       \multicolumn{5}{c}{BR1202-0725} \\
      \hline
      Quasar & [C\,{\sc ii}] & $4.6942$ & $\leq 4.3$ & $\leq 4.7$ \\
      SMG$^{\textrm{\dag}}$ & [C\,{\sc ii}] & $4.6915$ & $\leq 0.3$ & $\leq 0.5$ \\
      Ly$\alpha1^{\textrm{\dag}}$ & [C\,{\sc ii}] & $4.6950$ & $\leq 0.4$ & $\leq 0.7$ \\
      Ly$\alpha1^{\textrm{\dag}}$ & Ly$\alpha$ & $4.7026$ & $\leq 1.2$ & $\leq 0.3$ \\
      Ly$\alpha2^{\textrm{\dag}}$ & [C\,{\sc ii}] & $4.7055$ & $\leq 0.5$ & $\leq 0.6$ \\
      \hline
       \multicolumn{5}{c}{${\textrm{J1000+0234}}^{\textrm{\dag}}$} \\
      \hline
      SMG & Ly$\alpha$ & $4.547$ & $\leq 1.5$ & $\leq 1.3$ \\
      SMG & $^{12}$CO($4-3$) & $4.5423$ & $\leq 2.7$ & $\leq 1.1$ \\
      \hline
   \end{tabular}
    \begin{list}{}
      \centering
      \item $^1$ in units of $\times 10^{-17}\,$\ergscm.
      \item \dag \, flux limits are 5$\sigma$ level.
    \end{list}
  \label{redshift}
\end{table*}

\subsection{J1000+0234} 

\begin{figure*}
\centering
\includegraphics[width=\textwidth]{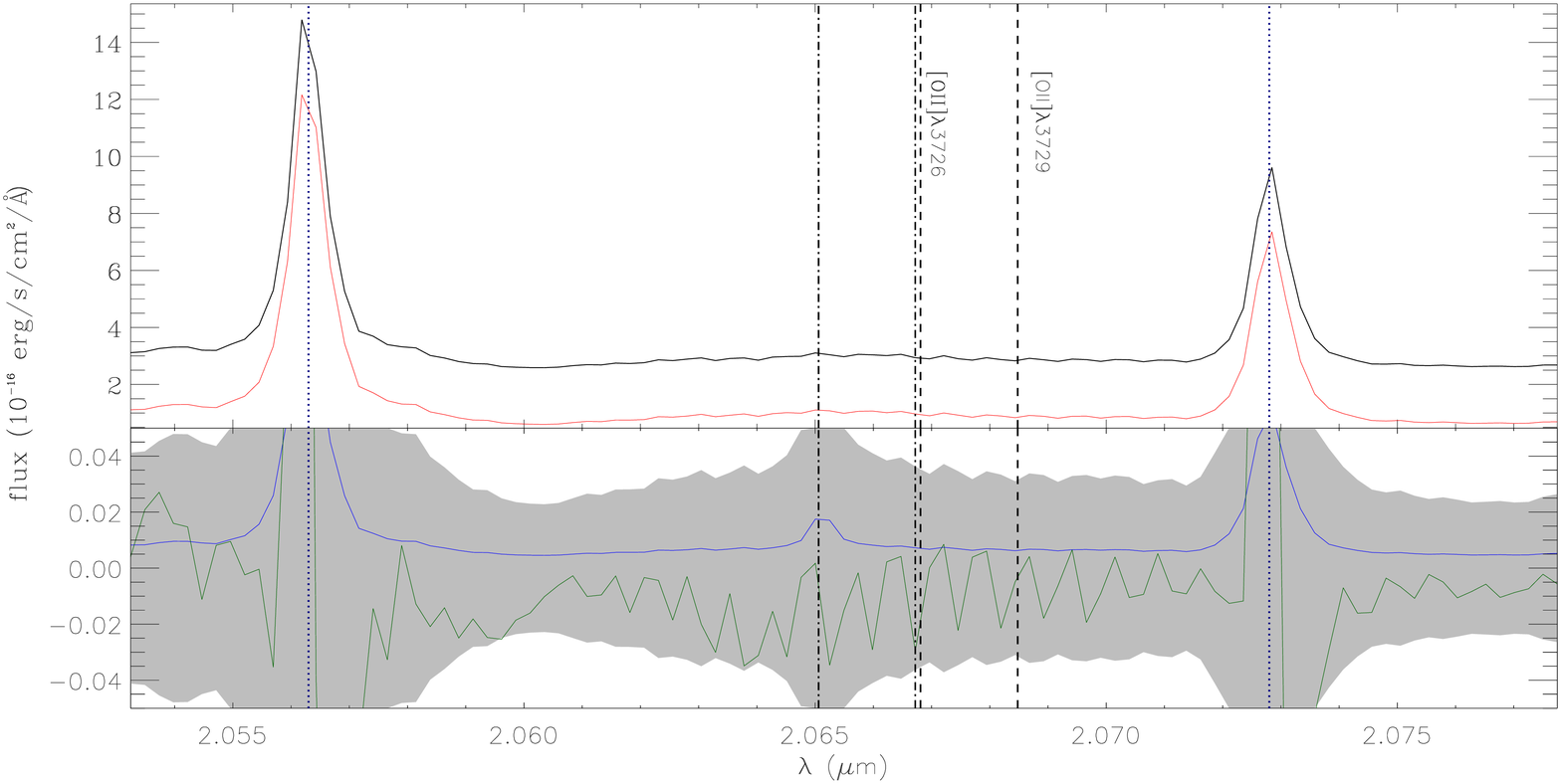}
\caption{Black line: Integrated spectrum of J1000+0234 SMG, extracted inside the white squares shown in right panel of Fig. \ref{hst}, added by a constant ($2 \times 10^{-16}$ \ergscma), for visual purposes. Red line: Sky spectra, which were derived as the median of the sky spectrum extracted from six different extraction regions located outside the overall structure of J1000+0234. Green line shows the residual between object and sky spectra and 1$\sigma$ uncertainty of the sky spectrum is displayed in blue line. Gray shaded area displays the 5$\sigma$ level. [O\,{\sc ii}] doublet wavelengths using redshifts from Ly$\alpha$ and $^{12}$CO($4-3$) emission-lines are represented by dashed and dot-dashed vertical black lines, respectively \citep[][see Table \ref{redshift}]{capak08,schinnerer08}. Dotted vertical lines correspond to sky lines wavelengths.}
\label{spec_j10}
\end{figure*}

The near-infrared structure of J1000+0234 is presented in right panel of Fig. \ref{hst} in a F160W {\it HST-WFC3} image, centered on the brightest region and covering the same FoV as our SINFONI integral field data. Fainter structures to the south-west and north-west appear within 1-2 arcsec. A foreground object ($z = 1.41$) is observed to the west of the SMG, as noted by \citet[][white arrow]{capak08}. White squares correspond to $0\farcs625 \times 0\farcs625$ apertures, that match our observational seeing (FWHM), where spectra were extracted. To optimize the signal from the SMG, the fluxes from the two regions identified as SMG have been combined. These are the brightest regions identified in the {\it HST-WFC3} image and where the emission from the [O\,{\sc ii}] would be expected in the {\it K}-band. Since the final SINFONI datacube was built using object and sky expositions, we establish the flux due to the sky emission (and its uncertainty) as the median value of the spaxels in six different extraction regions located outside the overall structure of J1000+0234 and avoiding other possible related features in the field. These regions have the same aperture size as the white squares in Fig. \ref{hst}.

The combined integrated spectrum of J1000+0234, extracted from the two apertures identified as SMG in Fig. \ref{hst}, is displayed in Fig. \ref{spec_j10}. Again there is no evidence of [O\,{\sc ii}] emission-lines, with no emission above the (5$\sigma$) sky uncertainties, and the sky lines are the only clear features present. The upper limits for the flux of the [O\,{\sc ii}] doublet is in the $\sim 2.8-3.8 \times 10^{-17}$ \ergscm range, depending on the redshift and line used (see Table \ref{redshift} for details).

\section{Discussions}
\label{analys}

\subsection{Star formation rates}
\label{SFRs}

We estimated the SFRs following the relation from \citet{kennicutt98}:

\begin{equation}
\textrm{SFR} (M_\odot\, yr^{-1}) = (1.4 \pm 0.4) \times 10^{-41} L_{\textrm{[O\,{\sc ii}] }} (\textrm{erg s}^{-1}),\end{equation}

\noindent
which make use of a Salpeter Initial Mass Function (IMF) with solar abundances. Table \ref{sfr} displays the upper limits of $L_{\textrm{[O\,{\sc ii}] }}$ and SFRs uncorrected from extinction for the objects of the two SMG systems, considering the redshifts mentioned in Table \ref{redshift} (lines used to measure the redshift are in parenthesis). Two approaches are shown: one using both of the [O{\sc ii}] emission-lines to estimate its luminosity, and another using the emission line less affected by the sky lines (thus with lower flux) to estimate the other, considering a line ratio of [O{\sc ii}]$3729/3726 = 1$ (assuming a $T \approx 15000\,$K and log\,$n_e = 2.5$) \footnote{SFRs should be multiplied by $0.6-0.7$ if applying another IMF such as \citet{kroupa01} or \citet{chabrier05} \citep{madau14}.}.

We find upper limits for the SFR rates of about $10-15\,$\myr for the different components of the BR 1202-0725 system, and in the $\sim 30-40\,$\myr range for J1000+0234. For the two SMGs, these SFR values are about two orders of magnitude lower than those derived from previous measurements, including mid-, far-infrared, and radio tracers. This clearly indicates that the amount of unobscured star formation in these systems as traced by the [O\,{\sc ii}] in emission is extremely low as compared with the global star formation. We find, for Ly$\alpha2$, SFRs of about one order of magnitude lower than those derived from \citet{carilli13}, using the FIR luminosity and the [C\,{\sc ii}] $158\mu$m line emission measurements. However, for Ly$\alpha1$, we find SFR upper limits that are in agreement with most of the estimates based on infrared tracers (see Table \ref{sfr}), indicating therefore that this system is less affected by dust internal extinction and is likely to have intrinsically low visual extinction values. 

Assuming internal extincion plays the main role in explaining the differences between our [O\,{\sc ii}]-derived SFR and those derived from the infrared, we estimate in the following section the average visual extinction in the two SMGs and the two LAEs of BR1202-0725.

\begin{table*}
   \centering   
   \caption{\it Derived upper limits for the [O\,{\sc ii}] luminosities and corresponding SFR.}
   \begin{tabular}{c c c c c c} 
      \hline \hline
      \multicolumn{6}{c}{BR1202-0725} \\
      \hline
      Object				& $\textrm{L}_{\textrm{[O{\sc ii}]}}^1$	& SFR$^1$ (this work)	& $\textrm{L}_{\textrm{[O{\sc ii}]}}^2$	& SFR$^2$ (this work)	& SFR$^3$ (literature) \\
					& [$\times 10^{42}\, \textrm{erg/s}$]		& [\myr]		& [$\times 10^{42}\, \textrm{erg/s}$]		& [\myr]		& [\myr] \\
      \hline
      SMG				& $\leq 1.8$					& $\leq 26$		& $\leq 0.7$					& $\leq 10$		& (1) \\
      Ly$\alpha1$ ([C\,{\sc ii}])	& $\leq 2.5$					& $\leq 34$		& $\leq 1.0$					& $\leq 14$		& (2) \\
      Ly$\alpha1$ (Ly$\alpha$)	& $\leq 3.4$					& $\leq 48$		& $\leq 0.7$					& $\leq 9$		& (2) \\
      Ly$\alpha2$			& $\leq 2.5$					& $\leq 35$		& $\leq 1.2$					& $\leq 16$		& (3) \\
      \hline
      \multicolumn{6}{c}{J1000+0234} \\
      \hline
      SMG (Ly$\alpha$)			& $\leq 5.8$					& $\leq 82$		& $\leq 2.8$					& $\leq 39$		& (4) \\
      SMG ($^{12}$CO($4-3$))		& $\leq 7.9$					& $\leq 111$		& $\leq 2.3$					& $\leq 32$		& (4) \\
     \hline
    \end{tabular}
    \begin{list}{}
      \centering
      \item $^1$ calculated using both [O{\sc ii}] emission lines; $^2$ calculated using the [O{\sc ii}] emission line less affected by sky emission lines and assuming a line ratio of [O{\sc ii}]$3729/3726 = 1$; $^3$ in parenthesis the emission-line or spectrum continuum used to calculate the SFR. All works use a Salpeter IMF with solar abundances.
      \item (1) [a] $2000$ (FIR); [b] $2600$ (FIR, SED model); [b] $6800\pm1000$ ([C\,{\sc ii}]); [b] $4000$ (radio, SED model).
      \item (2) [c] $15-54$ (UV); [d] $\approx 13$ (UV); [e] $\leq 36$ (FIR, SED model); [e] $19$ ([C\,{\sc ii}]).
      \item (3) [e] $170$ (FIR, SED model); [e] $70$ ([C\,{\sc ii}]).
      \item (4) [f] $1000-4000$ (FIR, SED model); [f] $3700\pm700$ (radio); [f] $2900\pm100$ ($3.6\mu$m band).
      \item Refs: [a] \citet{salome12}, [b] \citet{carniani13}, [c] \citet{fontana98}, [d] \citet{ohyama04}, [e] \citet{carilli13}, [f] \citet{capak08}
    \end{list}
    \label{sfr}
\end{table*}

\subsection{Internal extinction within the systems}
\label{reddening}

As already mentioned in the previous section, the most probable reason for the non detection of the [O{\sc ii}] emission-lines and estimated low SFR upper limits when compared with previous SFR values is extinction due to dust absorption in the SMGs. SFRs of $2000-7000\,$\myr and $1000-4000\,$\myr for BR1202-0725 and J1000+0234 SMGs would translate into $L_{\textrm{[O\,{\sc ii}]}}$ of $1.4-5.0 \times 10^{44}$ \ergs and $0.7-2.9 \times 10^{44}$ \ergs, respectively, using the same \citet{kennicutt98} relation. Using the upper limits for the luminosities we estimate from the [O\,{\sc ii}] emission line less affected by the sky (fourth column in Table \ref{sfr}), the ratio of the [O\,{\sc ii}]-derived to IR-derived [O\,{\sc ii}] luminosities imply an extinction of at least $A_{\textrm{[O{\sc ii}]}} \ge 5.7-7.1$ mag and $3.5-5.2$ mag, for BR1202-0725 and J1000+0234 SMGs respectively. The resulting visual internal extinction is $A_V \ge 3.4 - 4.3$ mag and $2.1 - 3.1$ mag for a SMC reddening law \citep{gordon03}, or $A_V \ge 3.9 - 4.9$ and $2.4 - 3.6$ mag for a starburst reddening law \citep{calzetti00}, for BR1202-0725 and J1000+0234 SMGs respectively. These extinction values are comparable with the ones found in a low-z sample of U/LIRGS, measured over sizes of a few to several kpc \citep{piqueras13}, in agreement with other studies that find evidence that these objects are local conterparts of high-$z$ SMGs \citep{nesvadba07,menendez13}.

For the cases of the two LAEs of the BR1202-0725 system, Ly$\alpha1$ and Ly$\alpha2$, previous SFR values cover the $13-54\,$\myr and $70-170\,$\myr ranges (see Table \ref{sfr}), corresponding to luminosity upper limits of $L_{\textrm{[O{\sc ii}]}} \approx 0.9-3.8 \times 10^{42}$ \ergs and $5.0 - 12.1 \times 10^{42}$ \ergs, respectively. The internal extincion in the visual is $A_V {\textrm{(SMC)}} \ge 0.2 - 0.9$ mag and $0.9 - 1.5$ mag, or $A_V {\textrm{(starburst)}} \ge 0.2 - 1.0$ mag and $1.1 - 1.7$ mag, for Ly$\alpha1$ and Ly$\alpha2$ respectively. It is important to emphasize here that the values of extinction in magnitudes we obtain are lower limits, since we derived them from flux upper limits, as show in Table \ref{redshift}.

\subsection{BR1202-0725: [O\,{\sc ii}] emission associated to Ly$\alpha$1?}
\label{ohta}

\citet{ohta00} reported [O\,{\sc ii}]$3727$ emission in a north-west companion of the quasar in BR1202-0725. This companion is located $\approx 2\farcs4$ northwest of the quasar, so probably relates to the object labeled as Ly$\alpha1$ here. Using a Cassegrain infrared camera (CISCO) attached to the Subaru Telescope, they obtained narrow-band imaging with a filter centered at $2.1196\,\mu$m with a FWHM of $0.0199\,\mu$m (the authors report a transmission between $2.113\,\mu$m and $2.126\,\mu$m ranging from 80\% to 88\%). The authors estimate a line flux of $2.5 \times 10^{-17}\,$\ergscm within a circular aperture of $0\farcs93$ radius, with a continuum level 10 times lower. 

According to our results, i.e. upper limit of 0.8 to 1.5 $\times$ 10$^{-17}\,$\ergscm for the [O\,{\sc ii}] emission, we hypothesize that the flux reported by Ohta and collaborators could likely be the result of a non-optimal sky subtraction, producing a positive residual interpreted as [O\,{\sc ii}] in emission. To test this hypothesis, we took an square aperture with a $0\farcs75$ on a side, centered on Ly$\alpha1$, and integrated the residual spectrum over the entire spectral window from $2.113\,\mu$m to $2.126\,\mu$m. We obtain an integrated flux of $2.9\pm0.2 \times 10^{-17}\,$\ergscm, in fair agreement with the value reported by \citet{ohta00}. The same kind of measurements were repeated at different positions within the SINFONI field-of-view, getting similar flux levels, within the $5 \sigma$ uncertainties. This supports the hypothesis of residual sky lines as the source of spurious emission. The residuals are likely due to two strong sky emission-lines, at $\approx \lambda 2.1177\,$ and $\approx \lambda 2.125 \,\mu$m (and another one, fainter, at $\approx \lambda 2.1232 \,\mu$m), that lay within the filter passband (Fig. \ref{spec_br}). 

\section{Conclusions}
\label{conc}

In this work we presented deep SINFONI {\it K}-band integral field spectroscopy of two SMG systems at $z \sim 5$, BR1202-0725 and J1000+0234. BR1202-0725 consists of the SMG, a QSO and two LAEs while J1000+0234 appears to have a complex extended structure as traced by multiband optical and near-infrared imaging. Our main conclusions are:

\begin{enumerate}
\item The spectra extracted for all the objects, including the two LAEs, do not show any signature of the [O\,{\sc ii}] doublet, within the 5$\sigma$ sky uncertainties. The corresponding upper limits for the SFR in the two SMGs are two orders of magnitude below those derived from far-infrared measurements. The differences are explained as due to internal obscuration equivalent to an average visual internal extinction of, at least, $4.1$ mag and $2.8$ mag for BR1202-0725 and J1000+0234, respectively. These average high internal extinctions are similar to those measured in low-z U/LIRGs;
\item The SFR upper limit for Ly$\alpha2$ derived from the non-detection of the [O\,{\sc ii}] is at least one order of magnitude lower than that measured from the infrared, while the corresponding values for Ly$\alpha1$ are in fairly good agreement, suggesting a very low internal extinction in this LAE. We find an internal extinction of at least $0.6$ and $1.3$ magnitudes in the visual for Ly$\alpha1$ and Ly$\alpha2$, respectively;
\item Previous claims of [O\,{\sc ii}] emission associated to BR1202-0725 Ly$\alpha1$ based on narrow-band imaging are not confirmed by our SINFONI 2D spectroscopy. Residuals due to some of the sky emission lines within the filter bandpass could produce fluxes compatible with the [O\,{\sc ii}] flux reported previously by \citet{ohta00}.
\end{enumerate}

\begin{acknowledgements}
GSC and LC acknowledge support from CNPq special visitor fellowship PVE 313945/2013-6 under the Brazilian program Science without Borders. LC acknowledge support by grant AYA2012-32295. JPL acknowledge support by grant AYA-2012-39408-C02-1. LC acknowledge support by grant AYA2012-39408-C02-01 and AYA2015-68964. Based on observations collected at the European Organisation for Astronomical Research in the Southern Hemisphere, Chile, programm 093.A-0204A. Some of the data presented in this paper were obtained from the Multimission Archive at the Space Telescope Science Institute (MAST). GSC wants to thank Natacha Z. Dametto for fruitful discussions and assistance.
\end{acknowledgements}

\bibliographystyle{aa} 
\bibliography{SMGs_accepted} 

\end{document}